\documentclass[12pt,utf8]{iopart}
\usepackage{booktabs}
\usepackage[pdftex]{hyperref,color,graphicx}
\hypersetup{pdftitle={Microwave control of atomic motional states in a spin-dependent optical lattice},pdfauthor={Andrea Alberti},colorlinks=true,linkcolor=blue,citecolor=blue,filecolor=blue,urlcolor=blue}

\makeatletter

\usepackage{iopams}
\usepackage{setstack}

\usepackage[utf8]{inputenc}
\usepackage[T1]{fontenc}
\usepackage[english]{babel}
\usepackage{siunitx}
\usepackage{braket}
\usepackage{subfig}

\newcommand{\figref}[2]{(\hyperref[#1]{\ref{#1}#2})}

\makeatother

\begin{document}

\title{Microwave control of atomic motional states in a spin-dependent optical lattice}

\author{Noomen Belmechri$^1$, Leonid Förster$^1$, Wolfgang Alt$^1$, 
Artur Widera$^2$, Dieter Meschede$^1$, and Andrea Alberti$^1$}

\address{$^1$ Institut für angewandte Physik, Universit\"at Bonn, Germany\\
$^2$ Fachbereich Physik und Forschungszentrum OPTIMAS, Universit\"at Kaiserslautern, Erwin-Schr\"odinger-Stra\ss{}e, D67663 Kaiserslautern, Germany}

\ead{alberti@iap.uni-bonn.de}
\begin{abstract}
Spin-dependent optical potentials allow us to use microwave radiation to
manipulate the motional state of trapped neutral atoms (\href{http://dx.doi.org/10.1103/PhysRevLett.103.233001}{Förster et al. 2009 \emph{\mbox{Phys. Rev. 
Lett.}} \textbf{103}, 233001}). Here, we discuss this method in greater detail, comparing it to the
widely-employed Raman sideband coupling method. We provide a simplified model for sideband
cooling in a spin-dependent potential, and we discuss it in terms of the generalized
Lamb-Dicke parameter. Using a master equation formalism, we present a quantitative analysis of
the cooling performance for our experiment, which can be generalized to other experimental
settings. We additionally use microwave sideband transitions to engineer motional Fock states
and coherent states, and we devise a technique for measuring the population distribution of the
prepared states.
\end{abstract}

\submitto{\jpb}
\maketitle

\tableofcontents{}

\section{Introduction}

Motional state control of atomic particles is achieved by the absorption
and emission cycles of a resonant or near resonant radiation, i.e.,
by light scattering typically at optical frequencies. For instance,
laser Doppler cooling reduces the momentum of atoms or ions
through multiple recoil processes \cite{wineland1}. Coherent momentum
transfer can be performed with two-photon Raman processes \cite{Chu2}
for applications in, e.g., atom interferometry \cite{interferometry}.

The quantum state of the atomic particles is composed of the internal states, e.g., two spin
states $\{\Ket{\uparrow},\Ket{\downarrow}\}$ for a two-level atom, and the external motional
state. For free particles the simplest motional state is the momentum state $|\vec{p\,}\rangle$.
Trapped particles are instead characterized by vibrational eigenstates $|n\rangle$, which in
the simplest case of a harmonic oscillator of frequency $\omega_{\text{vib}}$ have their
energies equally spaced as $\hbar\omega_{\text{{vib}}}(n+1/2)$.

In free space, the momentum state of a particle, and consequently its kinetic energy, is
changed by the momentum transfer $|\vec{p\,}\rangle\to|\vec{p}\,'\rangle$ in the
absorption/emission cycle of an optical photon. While the momentum transfer
picture also applies approx\-imately for trapped particles when the energy separation between  motional states is not spectroscopically resolved, 
recoil-free transitions become
possible in the resolved-sideband regime (M\"ossbauer effect). While carrier transitions do not change the vibrational quantum
state $\ket{n}$, the motional state can be controlled via sideband transitions
$\Ket{n}\to\Ket{n'}$ ($n'\neq n$), for instance, in incoherent cooling processes $\Ket{n}\to\Ket{n-1}$ or in coherent manipulation of vibrational states
\cite{Leibfried2003}. With trapped ions or neutral atoms trapped in optical lattices, the resolved-sideband regime is typically realized by two-photon Raman transitions connecting two different hyperfine ground states \cite{wineland,jessen,Perrin1998}. Alternatively,  in spin-dependent potentials it becomes also possible to use microwave transitions, which also offer sufficient spectral resolution \cite{Wunderlich2,durr,Weiss1997,Leonid,Ospelkaus}.

\begin{figure}[!b]
\begin{centering}
\includegraphics[scale=0.5]{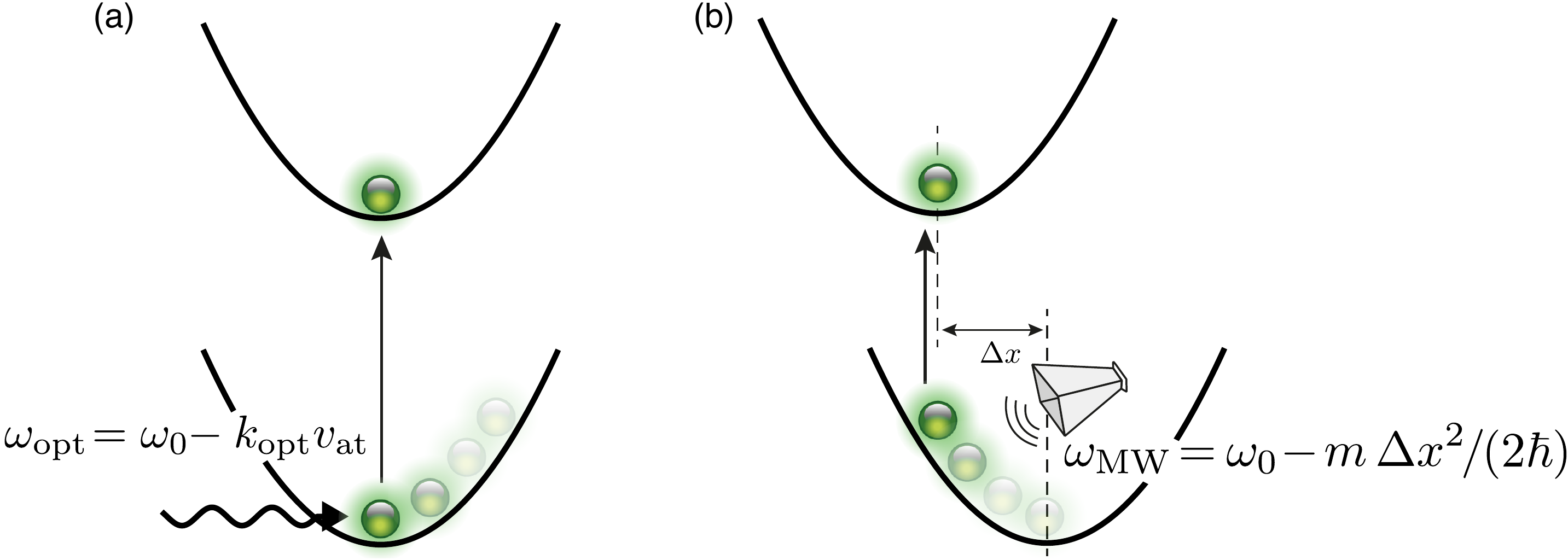}
\par\end{centering}

\caption{In a semi-classical picture, an atomic transition
can affect the motional state of an atom either (a) by a kinetic energy change caused
by the momentum transfer from an optical photon of wavevector $k_{\text{opt}}$  (velocity selective transition \cite{wineland1}), or (b) by a potential energy change when the potentials of the two internal states are displaced in space by $\Delta x$ (position selective transition). In the two cases, the motional energy is decreased when the detuning is set to the Doppler shift $k_{\text{opt}}v_{\text{at}}$ or the potential energy $m\,\omega^2\Delta x^2/(2\hbar)$, respectively. \label{fig:classicalProj}}
\end{figure}

In the semi-classical picture shown in figure (\ref{fig:classicalProj}),
an atomic transition exchanges either kinetic or potential energy
with the motional degree of freedom of the atom. With the absorption
of an optical photon, the kinetic energy is changed by the
momentum kick from the photon, and quantum mechanically the process
can be interpreted in terms of a displacement of the wavefunction
in momentum space. With the absorption of a microwave photon, which carries a
negligible momentum, the potential energy 
of the atom can be changed if the trapping potential
of the two states are different; this allows an interpretation in terms of a wavefunction displaced in position space.

% Removal of kinetic energy is routinely achieved by resolved sideband
% cooling as well as pulsed and continuous Raman cooling schemes applied
% to trapped ions or atoms. It is well known that in the Lamb-Dicke
% limit $\eta_{\Delta k}\ll1$ \cite{LambDickeLimit} the sideband coupling
% strengths of vibrational levels with quantum numbers differing by
% $\Delta n$ are proportional to $(\eta_{{\rm \Delta k}})^{n}$, with
% $\eta_{{\rm \Delta k}}=\hbar\Delta k/p_{0}$, and $\Delta k$ being
% the recoil transferred to the atom, and $p_{0}$ is the momentum-space
% extension of the lowest oscillator state wavefunction, yielding typically
% small values $\eta_{{\rm \Delta k}}$=0.01-0.1. 

% The coupling strength of the transition are described by a generalized
% complex Lamb-Dicke parameter $\eta$	 which can be continuously tuned
% via the wave packet spatial shift, typically to all values in the range
% $0\leqslant|\eta|\lesssim5$. The potential energy transfer in a state-dependent
% optical lattice thus allows efficient motion state engineering with
% neutral atoms, which is described in detail in this article. 

\section{Microwave induced motional sideband transitions}

\subsection{Motional states in a state dependent lattice\label{sub:TheorySB}}

We consider a single atom with two spin states \{$\Ket{\uparrow}$,
$\Ket{\downarrow}$\} trapped in a one-dimensional optical lattice.
We will initially ignore the internal degree of freedom of the atom
and take the Hamiltonian governing its motion in the trap as given
by
\begin{equation}
\hat{H}_{\text{{ext}}}=\frac{\hat{p}^{2}}{2m}+\frac{U_{0}}{2}\cos^{2}(k_{\text{{L}}}\hat{x}),\label{eq:ham}
\end{equation}
with $U_{0}$ being the trap depth, $k_{\text{{L}}}=2\pi/\lambda_{\text{L}}$
being the wavenumber of the two counter propagating laser fields creating
the lattice, and $\hat{x}$, $\hat{p}$, the atom's position and momentum,
respectively.

The motional eigenstates of an atom in such a potential are the well-known
Bloch wavefunctions $\Ket{\Psi_{n,k}^{B}}$, where $n$ is the band
index ($n=0$ for the first band) and $k$ is the wavevector in the first Brillouin zone (BZ). In
the limit of deep lattice potentials that we are considering here, the
atoms remain localized for the time scales of the experiment and their
spatial state is best described by the maximally localized Wannier
state \cite{kohn} 
\begin{equation}
\Ket{n,r}=\frac{1}{\sqrt{N}}\sum_{k\in\text{{BZ}}}e^{-ikrd}\Ket{\Psi_{n,k}^{B}}.\label{eq:Wannier}
\end{equation}
Here $N$ is the lattice size, $r$ the site index, and $d=\lambda_{\text{L}}/2$
the lattice spacing. In this deep lattice regime, we can safely view
the vanishingly narrow energy bands $\varepsilon_{n}(k)$ as the vibrational
level energies $\varepsilon_{n}$ of the corresponding Wannier state
$\vert n,r\rangle$ at lattice site $r$; in the harmonic approximation
we would have $\varepsilon_{n}=\hbar\omega_{\text{vib}}(n+1/2)$.

\begin{figure}
\begin{centering}
\includegraphics[width=0.8\textwidth]{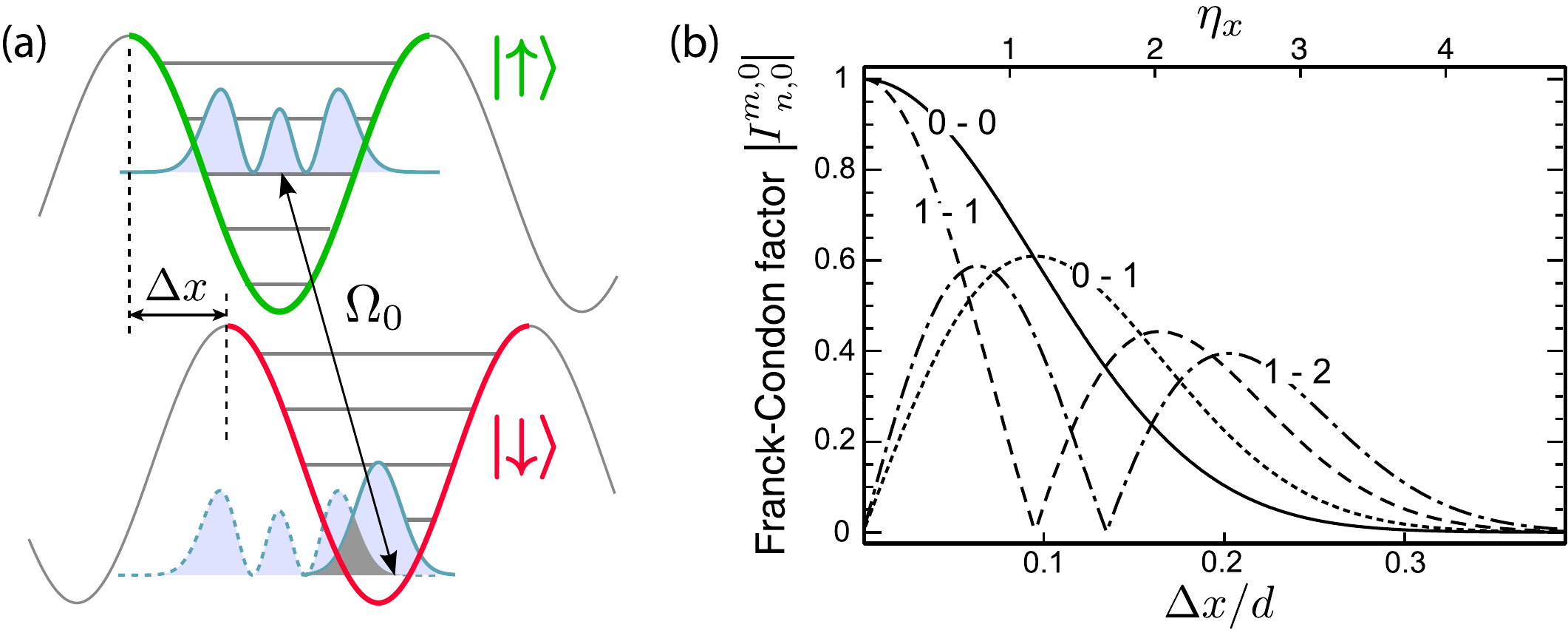}
\par\end{centering}
\centering{}\caption{(a) \label{fig:SDTImage+overlap}The coupling strength of a sideband
transition in a spin-dependent lattice is the bare spin state coupling
$\Omega_{0}$ multiplied by the overlap between the two involved
vibrational states, the Franck-Condon factor, which is controlled
by the relative shift $\Delta x$ between the two lattices. $\eta_{x}$
is the spatial Lamb-Dicke parameter defined in section \ref{sub:TheorySB} and later in \ref{sub:3.1}.
(b)\label{fig:FCfsCalc} Lattice shift dependence of the Franck-Condon
factors for different transitions, denoted as $n-m$, calculated
for typical experimental parameters (see text).\label{fig:Intro+FCfCalc}}
\end{figure}

The Wannier states form an orthonormal basis set such that the overlaps
between two different states yield $\Braket{n,r|n'\!,r'}=\delta_{n,n'}\delta_{r,r'}.$
This means that the interaction of the atomic spin with a microwave
field will fail to induce motional sideband transitions, $\Ket{n,r}\leftrightarrow\Ket{n'\!,r'}$,
because of the nearly negligible momentum carried by microwave photons,
five orders of magnitude smaller than that by optical photons. This
restriction can be lifted if the atom experiences a different trapping
potential depending on its internal spin state as the corresponding
motional eigenstates are then no longer orthogonal \cite{wunderlich2001,weiss2}.
A simple relative spatial shift of the potentials trapping each internal
state induces such a difference. A shift by a distance $\Delta x$
is accounted for by the position space shift operator $\hat{T}_{\Delta x}\equiv\exp(-i\:\hat{p}\Delta x/\hbar)$,
see figure \figref{fig:SDTImage+overlap}{a}. The overlap between the
two Wannier states then becomes 
\begin{equation}
\Braket{n'\!,r'|\hat{T}_{\Delta x}|n,r}\equiv I_{n,r}^{n'\!,r'}(\Delta x).\label{eq:Txoverlap}
\end{equation}
The resulting overlap integral, $-1\leq I_{n,r}^{n'\!,r'}(\Delta x)\leq1$,
is hence a known function of $\Delta x$, see figure \figref{fig:SDTImage+overlap}{b}.
It is analogous to the Franck\textendash{}Condon factor from molecular
physics and it determines the strength of the transitions coupling
different vibrational levels \cite{Bernath1995}.

One way to realize the shift operator $\hat{T}_{\Delta x}$ is by
two overlapped lattices which trap each spin state separately and
can be independently shifted in the longitudinal direction as shown
in figure (\ref{fig:SDTImage+overlap}). The trapping potential thus
becomes dependent on the spin state $s=\{\uparrow,\downarrow\}$ and
the shift distance $\Delta x=x_{\uparrow}^{0}-x_{\downarrow}^{0}$,
\begin{equation}
\hat{H}_{\text{{ext}}}=\frac{\hat{p}^{2}}{2m}+\sum_{s=\{\uparrow,\downarrow\}}\frac{U_{0}^{s}}{2}\cos^{2}[k_{\text{{L}}}(\hat{x}-x_{s}^{0})]\otimes\Ket{s}\Bra{s}\label{eq:newH}
\end{equation}
with $x_{s}^{0}$ being the position of the lattice trapping the state $\Ket{s}$.
The total transition matrix element for two spin states coupled by
an interaction Hamiltonian $H_{\text{{I}}}$, with a free-atom bare
Rabi frequency $\Omega_{0}$, is then given by 
\begin{equation}
\hbar\Omega_{n,r}^{n'\!,r'}(\Delta x)/2=\left\langle s'\!,n'\!,r'\left\vert \hat{T}_{\Delta x}\otimes H_{\text{{I}}}\right\vert s,n,r\right\rangle =I_{n,r}^{n'\!,r'}(\Delta x)\times\hbar\Omega_{0}/2.\label{eq:SBcoupling}
\end{equation}
The Franck-Condon factors $I_{n,r}^{n'\!,r'}(\Delta x)$ can be explicitly
evaluated using equations (\ref{eq:Wannier}) and (\ref{eq:Txoverlap}).
We first rewrite equation (\ref{eq:Wannier}) using Bloch's theorem,
\begin{equation}
W_{n,r}(x)=\frac{1}{\sqrt{N}}\sum_{k\in\text{{BZ}}}\:\sum_{q\in\mathbb{Z}}e^{-ikrd}e^{i\frac{2\pi}{d}q}\: a_{n,q}(k)\;\Ket{k}.\label{eq:BlochTheorem}
\end{equation}
with $a_{n,q}(k)$ being the Fourier coefficients of the Bloch functions
and $\Ket{k}$ the planewave state. These functions can be constructed
using the periodic solutions of the Mathieu differential equation
\cite{mathieu,slater} with their phase chosen such that the resulting
Wannier states are real and have the proper parity corresponding to
their respective vibrational levels \cite{kohn}. The coefficients
$a_{n,q}(k)$ are numerically obtained from algorithms for the computation
of Mathieu coefficients \cite{alhargan}. Inserting (\ref{eq:BlochTheorem})
in (\ref{eq:Txoverlap}) and taking into account the parity of the
Wannier states, or equivalently the parity of the band $n$, one eventually
arrives at the following expression for the Franck-Condon factors
\begin{equation}
I_{n,r}^{n'\!,r'}(\Delta x)=2\sum_{k\in\text{{BZ}}}\:\sum_{q\in\mathbb{Z}}\mathcal{F}\left[(k+\frac{2\pi}{d}q)(\Delta x+r-r')\right]\: a_{n,q}^{*}(k)\, a_{n'\!,q}(k),\label{eq:FCFexpression}
\end{equation}
where we have defined $\mathcal{F}(x):=\cos(x)\,$ if $n$ and $n'$
have the same parity, and $\mathcal{F}(x):=\sin(x)\,$ otherwise.
Numerical evaluation of (\ref{eq:FCFexpression}) is shown in figure
(\ref{fig:FCfsCalc}).

Considering a single lattice site and assuming the harmonic approximation for the potential, the shift operator takes the simple form $\hat{T}_{\Delta x}=\exp[\eta_x(a^\dagger -a)]$, where $a^\dagger$ ($a$) is the raising (lowering) operator acting on the vibrational states. Here we introduced the spatial Lamb-Dicke parameter 
\begin{equation}
\eta_x=\Delta x/(2\hspace{1pt}x_0)\,,
\end{equation}
where $x_0$ is equal to the rms width of the motional ground state. When $\eta_x\ll1$, taking the first order term in $\eta_x$ of $\hat{T}_{\Delta x}$ allows for a simple expression of the Franck-Condon factors for transitions on the same lattice site (i.e., $r$=$r'$=0), $I_{n,0}^{n'\!,0}(\Delta x)\approx\delta_{n,n'}+\eta_x(\sqrt{n'}\delta_{n'\!,n+1}-\sqrt{n}\delta_{n'\!,n-1})$. 

\subsection{Experimental setup\label{sub:sec2.2}}
We load Cesium ($^{133}\text{Cs}$) atoms from a magneto optical trap into a 1D optical lattice
formed by two counter-propagating, far-detuned, linearly polarized laser beams. The filling factor is at most one atom per lattice site due to light-induced collisions \cite{Schlosser:2001}. A weak guiding magnetic
field of 3\;G oriented along the lattice lifts the degeneracy between the Zeeman sublevels of
the Cesium $6^{2}S_{1/2}$ ground state such that atoms can be initialized
by optical pumping beams into the hyperfine state $\Ket{\uparrow}\equiv\Ket{F=4,m_{F}=4}$.
Microwave radiation, at around $\omega_\text{MW}=2\pi\times\SI{9.2}{\giga\hertz}$, couples states $\Ket{\uparrow}$
and $\Ket{\downarrow}\equiv\Ket{F=3,m_{F}=3}$ with the bare Rabi frequency
of $\Omega_{0}=2\pi\times\SI{60}{\kilo\hertz}$ \cite{addressing}. The
spin state of the atom is probed using the so called ``push-out''
technique \cite{pushout} which consists of counting the fraction
of atoms left in $\Ket{\downarrow}$ after all the atoms in $\Ket{\uparrow}$
have been removed by an intense radiation pulse.

An angle $\theta$ between the linear polarization vectors of the two beams
forming the lattice is equivalent in the circular basis to a phase
delay of $2\theta$ between two collinear and independent circularly-polarized
standing waves, $\sigma^{+}$ and $\sigma^{-}$, or equivalently
to a standing wave longitudinal relative shift of 
\begin{equation}
\Delta x_\text{sw}(\theta)=\theta\, d/\pi.\label{eq:shiftdist}
\end{equation}
The polarization angle $\theta$ is controlled by an electro-optical
modulator (EOM) and two quarter-wave plates in the path of one of
the two lattice beams. The two in-phase circular components of the
beam are mapped by the first $\lambda_{\text{L}}/4$ plate onto orthogonal linear
polarizations parallel to the EOM axes. The retardation
$2\theta$ induced by the EOM is proportional to the voltage signal applied to it. The last plate then
converts the linear polarizations back into the circular ones while
conserving the delay.

\begin{figure}[t]
\begin{centering}
\includegraphics[scale=0.45]{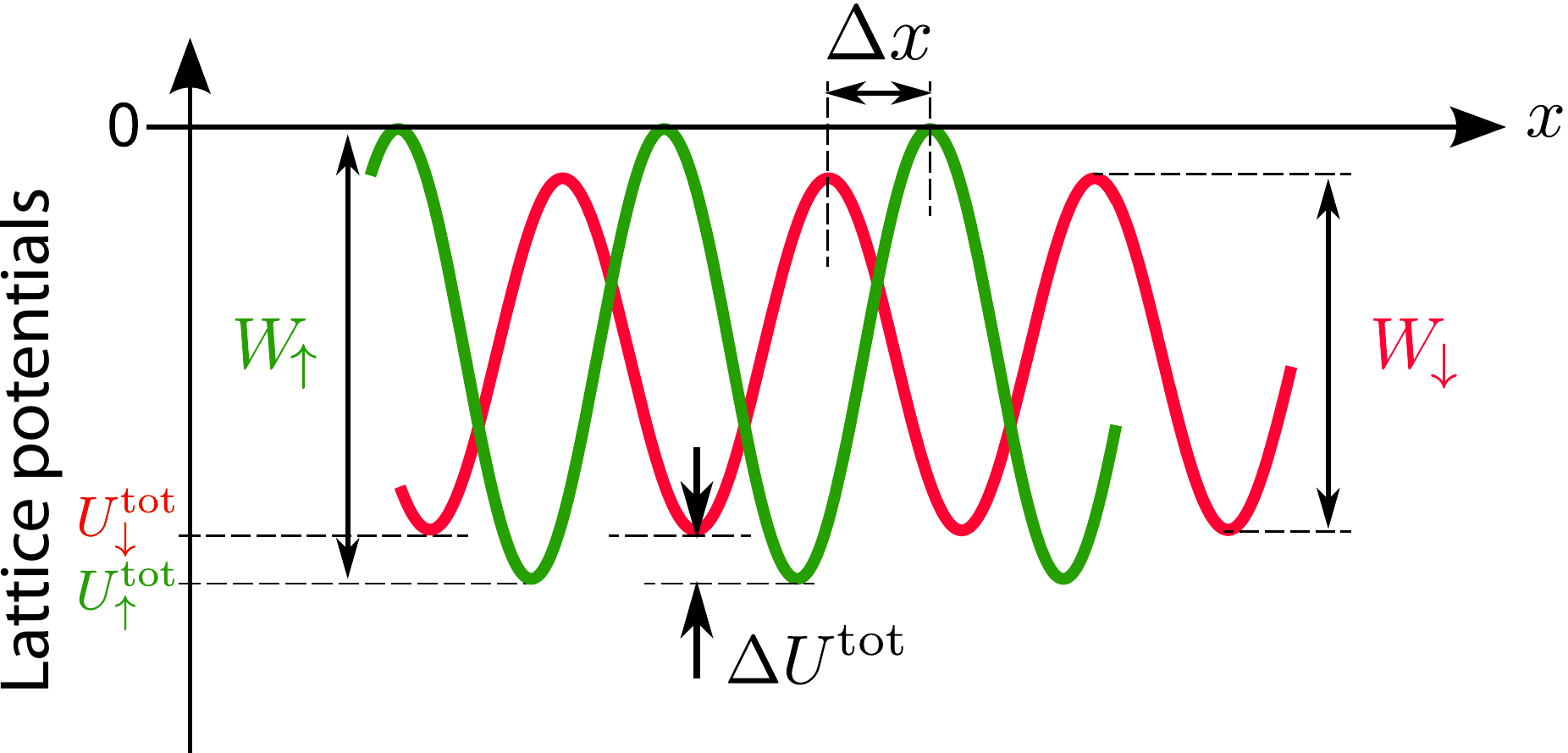}
\par\end{centering}

\caption{State-dependent optical lattices relatively shifted by a distance
$\Delta x$. The total trap depth difference $\Delta U^{\text{tot}}=U_{\uparrow}^{\text{tot}}-U_{\downarrow}^{\text{tot}} $, and lattice contrast
$W_s$ for spin state $\Ket{s}$ are shown. Unlike the spin $\Ket{\uparrow}$ lattice,
the contrast and total depth of the spin $\Ket{\downarrow}$ lattice
vary with the shift distance.\label{fig:contrast}}
\end{figure}

The trapping potentials resulting from the $\sigma^{+}$ and $\sigma^{-}$ standing waves for a spin state $\Ket{s}$ are
\begin{equation}
	U_{s}=U_{s}^{\textrm{\text{{tot}}}}+W_{s}\cos^{2}[k_{L}(x-x_{s}^{0})]\label{eq:latpotgen} \end{equation}
where $W_{s}$ and $U_{s}^{\text{{tot}}}$ are the lattice contrast, taking positive values, and total trap depth for state $\Ket{s}$, respectively. Both $W_{s}$ and $U_{s}^{\text{{tot}}}$ depend on the lattice lasers wavelength $\lambda_{\text{L}}$ and on the lattice shift $\Delta x$ or equivalently the polarization angle $\theta$, see figure (\ref{fig:contrast}). For alkali atoms, one can define the ``magic wavelength'' as the one where the state $\Ket{\uparrow}$ experiences the $\sigma^{+}$ standing wave only. This occurs at $\lambda_{\text{L}}=\lambda_2+(\lambda_1 - \lambda_2)/(2\lambda_1/\lambda_2+1)\approx \lambda_2+(\lambda_1 - \lambda_2)/3 $, where $\lambda_1$ ($\lambda_2$) is the wavelength of the D${}_1$ (D${}_2$) line \cite{Deutsch1998,Jaksch99}, which is $\lambda_{\text{L}}=\SI{866}{\nano\meter}$ in our case. At this wavelength, for the spin $\Ket{\uparrow}$ state, equation (\ref{eq:latpotgen}) reads
\begin{equation}
	U_{\uparrow}=-W_{\uparrow}+W_{\uparrow}\cos^2(k_{L}x-\theta/2),\label{eq:Uup}
\end{equation}
while the spin $\Ket{\downarrow}$ state experiences both $\sigma^{+}$ and $\sigma^{-}$ standing waves with a relative weight of $1/8$ and $7/8$, respectively. The lattice potential in this case is 
\begin{equation}
	U_{\downarrow}=-W_{\uparrow}+(1/8)\hspace{1pt}W_{\uparrow}\cos^{2}(k_{L}x-\theta/2)\;+\; (7/8)\hspace{1pt}W_{\uparrow}\cos^{2}(k_{L}x+\theta/2)\label{eq:Udown}\,.
\end{equation}
With the notation of equation~(\ref{eq:latpotgen}), one finds that $W_\uparrow=-U_\uparrow^\text{tot}$ is independent from the angle $\theta$, while $W_\downarrow=[\cos(\theta)^2+(3/4)^2\sin(\theta)^2]^{1/2}\hspace{2pt}W_\uparrow$ and $U_\downarrow^\text{tot}=-(W_\uparrow+W_\downarrow)/2$. In addition, one obtains the lattice relative shift $\Delta x=(d/\pi)\{\theta+\arctan[3\tan(\theta)/4]\}/2$.
Equations (\ref{eq:Uup}) and (\ref{eq:Udown}) constitute the closest realization of the idealized spin-dependent lattice discussed in section \ref{sub:TheorySB}. The small admixture of a $\sigma^{+}$ component in equation (\ref{eq:Udown}) results in a lattice depth $W_{\downarrow}$ that depends on $\theta$, or equivalently on the lattice shift $\Delta x$, which makes the energy levels $\varepsilon_{s,n}(\Delta x)$ depend on the spin state and on the shift $\Delta x$, see figure (\ref{fig:contrast}). The nonlinear position shift of the $U_\downarrow$ potential, $x_{\downarrow}^{0}$, makes $\Delta x$ deviate from the standing wave relative shift $\Delta x_\text{sw}$ in equation (\ref{eq:shiftdist}), and this has to be taken into account in the calculation of the Franck-Condon factors \cite{Deutsch1998}.

The typical total lattice depth used in our experiment is $W_\uparrow\approx850\, E_{\text{R}}^{\text{latt}}$ (corresponding to $\SI{80}{\micro\kelvin}$),
with $E_{\text{R}}^{\text{latt}}=\hbar^{2}k_{\text{L}}^{2}/2m_{\text{Cs}}$
as the lattice recoil, which amounts to an oscillation frequency along
the lattice axis of $\omega_{\text{{vib}}}\approx2\pi\times\SI{116}{\kilo\hertz}$.
In the transverse direction, atoms are confined only by the Gaussian
profile of the lattice lasers which results in a transverse oscillation
frequency of $\omega_{\text{{rad}}}\approx2\pi\times\SI{1}{\kilo\hertz}$.
The typical initial temperature of the atoms loaded into the lattice
is $T\approx\SI{10}{\micro\kelvin}$, which in the harmonic approximation
amounts to mean vibrational numbers of $\overline{n}_{\text{{vib}}}\approx1.4$
and $\overline{n}_{\text{{rad}}}\approx280$ in the axial and transverse
directions, respectively.

\subsection{Microwave sideband spectra\label{sub:2.3}}

\begin{figure}[b]
\begin{centering}
\hfill{}\includegraphics[width=11.7cm]{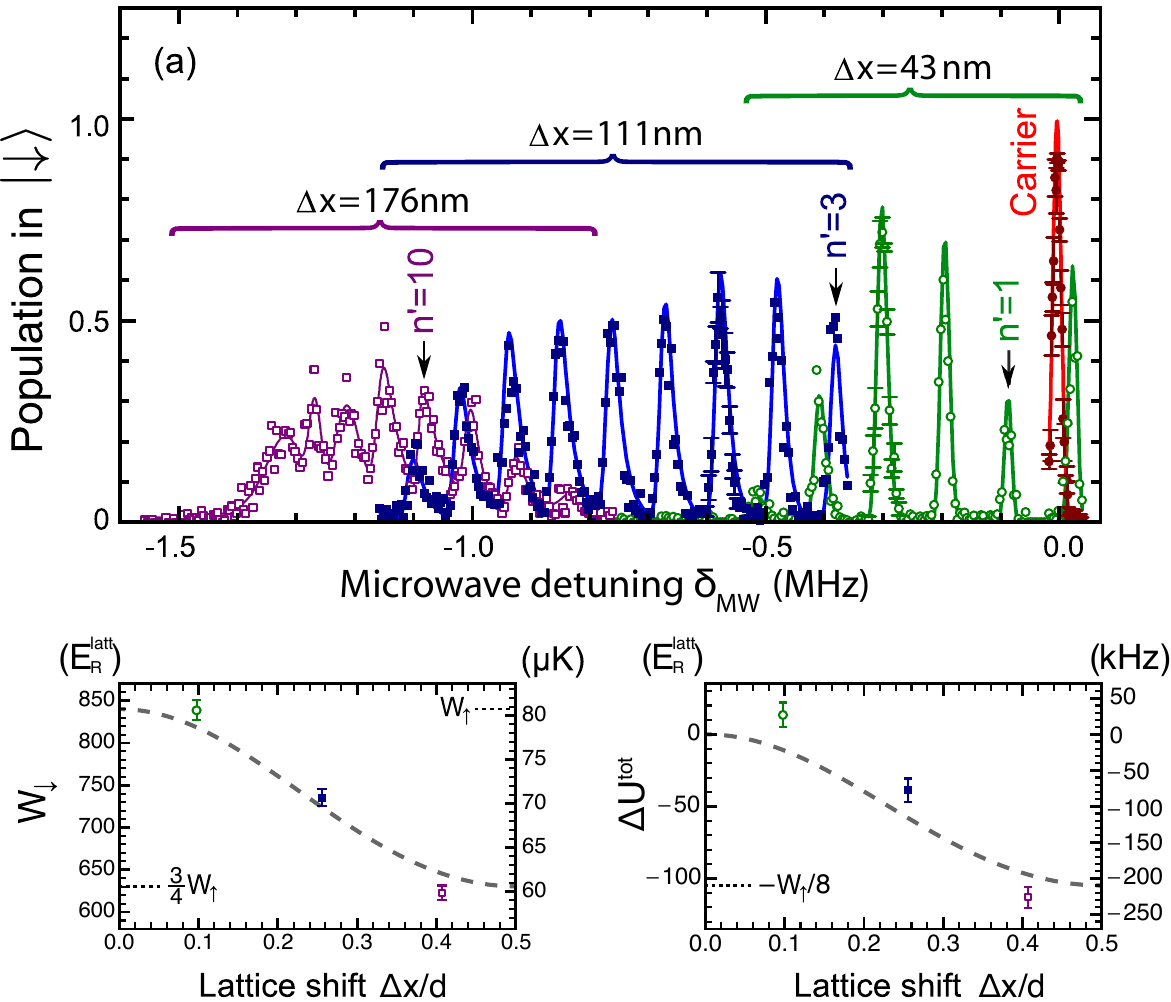}\hfill{}
\par\end{centering}

\centering{}\caption{(a) Microwave spectrum of sideband transitions $\Ket{\uparrow,n=0}\leftrightarrow\Ket{\downarrow,n'}$
for lattice shifts $\Delta x=$\{0nm ({\small$\bullet$}), 43nm ($\circ$),
111nm({\tiny$\blacksquare$}), 176nm ({\tiny$\square$})\} corresponding
to the parameter $\eta_{x}={\{0,1.2,3.1,4.9\}}$ defined in section \ref{sub:3.1} (data points from \cite{Leonid}). The microwave
detuning is given with respect to the carrier transition frequency.
Data points are the average on about 100 atoms and they are here fitted with a model that takes into account broadening
mechanisms detailed in the text. The error bars, reported only for three representative peaks, are obtained with the 68\% Clopper-Pearson interval method for binomial statistics. The panels (b) and (c) compare the expected
values (dashed lines) for the lattice contrast $W_{\downarrow}$ and total trap depth
difference $\Delta U^{\text{tot}}$ (see figure (\ref{fig:contrast}) and text in section~\ref{sub:sec2.2}) with the values extracted from the fits ($1\%$ uncertainty).\label{fig:fullSBspectrum}}
\end{figure}

We investigate sideband transitions by recording microwave spectra
for different lattice shifts. Controlling the relative distance $\Delta x$ allows us to continuously tune the parameter $\eta_x$ from 0 to about 5.
In order to resolve the sidebands we use Gaussian
microwave pulses with a FWHM of $\SI{30}{\micro\second}$ and a
bare Rabi frequency of $\Omega_{0}/2\pi=\SI{36}{\kilo\hertz}$, corresponding
to the $\pi$-pulse condition for the carrier transition. Figure (\ref{fig:fullSBspectrum})
shows a combined spectrum where transitions from $n=0$ to levels
up to $n'=14$ are well resolved \cite{Leonid}. Four spectra are
recorded for four different lattice shifts. With an unshifted lattice
only the carrier transition is visible, and it defines the zero of
the microwave detuning $\delta_\text{MW}$. The remaining three lattice shifts were chosen
such that for each shift distance $\Delta x$ the sideband coupling
strength on the same site (i.e., $r$=$r'$), $\Omega_{n,0}^{n'\!,0}(\Delta x)$, is simultaneously close to maximum
for a small group of adjacent sideband transitions. The coupling strength for sites $r\neq r'$ can be neglected at the given shifts.

For each shift distance $\Delta x$ the microwave spectra are fitted
using the spectra yielded by a numerical calculation of the time evolution based on the following Hamiltonian
\begin{equation}
\hat{H}=\hat{H}_{0}+\hat{H}_{\text{MW}}\label{eq:H}
\end{equation}
with
\begin{eqnarray}
\hat{H}_{0} & = & \sum_{s=\uparrow,\downarrow}\sum_{n}\bigg(\varepsilon_{s,n}(\Delta x)+\delta_{s,\uparrow}\,\hbar\hspace{0.5pt}\omega_\text{HS}\bigg)\Ket{s,n,r}\Bra{s,n,r}\label{eq:H0}\,,\\
\hat{H}_{\text{MW}} & = & -\frac{\hbar}{2}\Omega_{0}\sum_{r,r'}\sum_{n,n'}\, I_{n,r}^{n'\!,r'}(\Delta x)\,\bigg(e^{-i\hspace{0.3pt}\omega_{\text{MW}}t}\Ket{\uparrow,n,r}\Bra{\downarrow,n'\!,r'}+\text{h.c.}\bigg)\,,\label{eq:HMW}
\end{eqnarray}
where $\omega_\text{HS}$ denotes the hyperfine splitting frequency of the ground state. With this notation, the microwave detuning reads $\delta_\text{MW}=\omega_\text{MW}-\omega_\text{HS}$.

Given the deep lattice regime considered here, in the numerical solution
of equation (\ref{eq:H}) the maximum number of vibrational levels
per site can be restricted with a good approximation to $n_{\text{max}}=15$,
before atoms start to behave like free particles tunneling between
sites or directly coupling to the continuum. In this regime, the coupling
strength for a sideband transition between two lattice sites separated
by a distance $x>d$ are two orders of magnitude lower than the typical
time scales of our experiment; therefore, we limit the site indices to $r=r'$.

\begin{figure}
\begin{centering}
\hfill{}\includegraphics[width=0.74\textwidth]{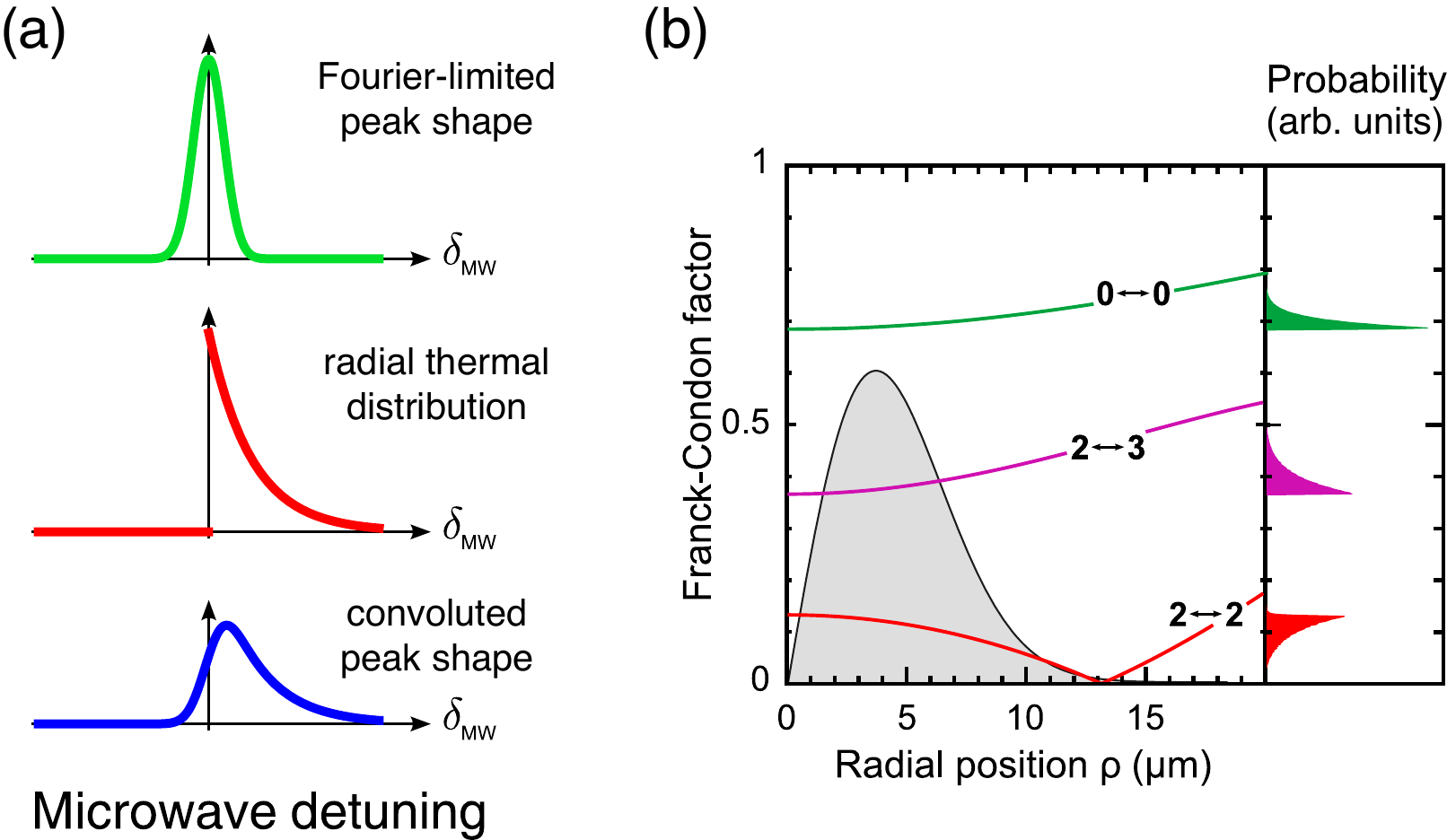}\hfill{}
\par\end{centering}

\centering{}\caption{(a) \label{fig:InhBroadeningConvol}\label{fig:FCFsDist}
Inhomogeneous broadening effect
due to the transverse motion of the atoms in the trap. The overall peak profile (bottom curve) is the convolution of the other two profiles. The Fourier-limited FWHM and the thermal broadening are typically $\SI{20}{\kilo\hertz}$ and $\SI{5}{\kilo\hertz}$ (for $T_\text{2D}\sim\SI{10}{\micro\kelvin}$ and first sideband), respectively. (b) Left panel:
Franck-Condon factors as a function of the radial distance $\rho$ (here, $\theta=\SI{15}{\degree}$). The gray profile shows the 2D radial distribution from equation~(\ref{eq:boltzmann}) for the same temperature. Right panel: Resulting thermal distribution of Franck-Condon factors.}
\end{figure}

In the fitting of the sideband spectra, the energy levels $\varepsilon_{s,n}$ and Franck-Condon
factors $I_{n,0}^{n'\!,0}$ depend on the fitting parameters $\Delta x$, $U_{s}^{\text{{tot}}}$
and $W_{s}$. In particular, in the harmonic approximation the spacing between two adjacent peaks is equal to the trap frequency of the $U_\downarrow$ potential, which therefore determines the lattice contrast $W_\downarrow$; the absolute offset of each spectrum is mainly determined by  the difference of the total trap depths, $\Delta U^{\text{{tot}}}=U_{\uparrow}^{\text{{tot}}}-U_{\downarrow}^{\text{{tot}}}$, expressed in frequency units. Additionally, an average over the thermal motion of the atoms in the transverse direction of the one-dimensional optical lattice has to be performed.
In fact, the lattice parameters $U_s^\text{tot}$ and $W_s$ depend on the transverse position
of the atom, and to take this dependence into account we assume that
during the microwave dynamics an atom has a ``frozen'' transverse
position $\rho$. This assumption is justified by the slow transverse
motion of the atoms, $\omega_{\text{rad}}/2\pi\approx \SI{1}{\kilo\hertz}$, compared
to the lowest bare Rabi frequency used for the microwave pulse, $\Omega_{0}/2\pi\approx\SI{14}{\kilo\hertz}$.
The transverse positions of the atoms are then assumed to be distributed
according to a two-dimensional Boltzmann distribution, shown in figure
\figref{fig:FCFsDist}{b} and given in the harmonic approximation by
\begin{equation}
\mathcal{P}(\rho)=\frac{\rho}{\sigma^{2}}\exp(-\frac{\rho^{2}}{2\sigma^{2}}),\qquad\text{ with }\qquad\sigma=\sqrt{\frac{k_{B}T_{\text{{2D}}}}{m_{\text{{Cs}}}\hspace{2pt}\omega_{\text{{rad}}}^{2}}}\label{eq:boltzmann}
\end{equation}
with $T_{\text{{2D}}}$ the transverse temperature. The thermal transverse
position distribution results in an inhomogeneous distribution of
microwave sideband resonance frequencies and of Franck-Condon factors, shown qualitatively in figure (\ref{fig:InhBroadeningConvol}). Both distributions are used to weight the
calculated spectra, with $T_{\text{{2D}}}$ as an additional fitting
parameter. The figure shows that thermal broadening effect becomes larger for higher sidebands, exhibiting a more pronounced asymmetric peak shapes. This behavior has a clear explanation: in the harmonic approximation, for instance, one expects the thermal broadening to increase linearly with band index $n$ while the Fourier-limited FWHM remains constant.

The best-fit results for $W_{\downarrow}$ and $\Delta U^{\text{{tot}}}$
are shown in figures \figref{fig:fullSBspectrum}{b} and \figref{fig:fullSBspectrum}{c}. This method allows us to spectroscopically determine the parameters of the spin-dependent potentials seen by the atoms with a relative uncertainty of about $1\%$. The small deviations from the expected values (dashed curves in the figure) can be attributed in part to measurement uncertainty and to polarization imperfections in the standing wave beams, resulting in slightly distorted potentials. For instance, polarization distortion can be responsible for the non monotonic behavior of the data points in figure~\figref{fig:fullSBspectrum}{c}.
From the fit, we obtain a temperature of $T_{\text{{2D}}}= (2.7\pm 0.5)\,\si{\micro\kelvin}$. Without axial ground state cooling, we measure a three-dimensional temperature of $\SI{10}{\micro\kelvin}$ by means of the adiabatic lowering technique~\cite{Wolfgang}. This discrepancy requires further investigations.

\section{Microwave sideband cooling}

The general principle of resolved sideband cooling, depicted in figure
(\ref{fig:ToyModel}), relies on the repetition of cooling cycles
where each cycle starts by a sideband transition $\Ket{\uparrow,n}\to\Ket{\downarrow,n-1}$
removing a vibrational energy quantum $\hbar\omega_{\text{{vib}}}$.
The cycle is then closed by an optical repumping process with a transition
to an optically excited state $\Ket{e}$ followed by a spontaneous
decay to the initial spin state. Because of the optical repumping,
the motional energy of the atom in each cycle increases on average,
which corresponds to heating. Therefore, in order to achieve cooling
the overall energy gained by an atom after one cycle must be negative.
In general, heating is caused by the momentum recoil from the optical
repumping photons, i.e. recoil heating. In the microwave-based scheme
however, shown in figure \figref{fig:ToyModel}{b}, an additional source
of heating, called hereafter ``projection heating,'' is present.
It is due to the difference between the trapping potentials of the
internal states, in this case it is a spatial shift. This difference
makes the projection of a vibrational state $\Ket{\downarrow,n}$
on an arbitrary state $\Ket{\uparrow,m}$ appreciable in contrast
to the case of identical potentials where transitions beyond $m=n,n\pm1$
are negligible.

\subsection{Raman vs.\ microwave sideband cooling\label{sub:3.1}}

\begin{figure}[!tb]
\begin{centering}
\includegraphics[scale=0.8]{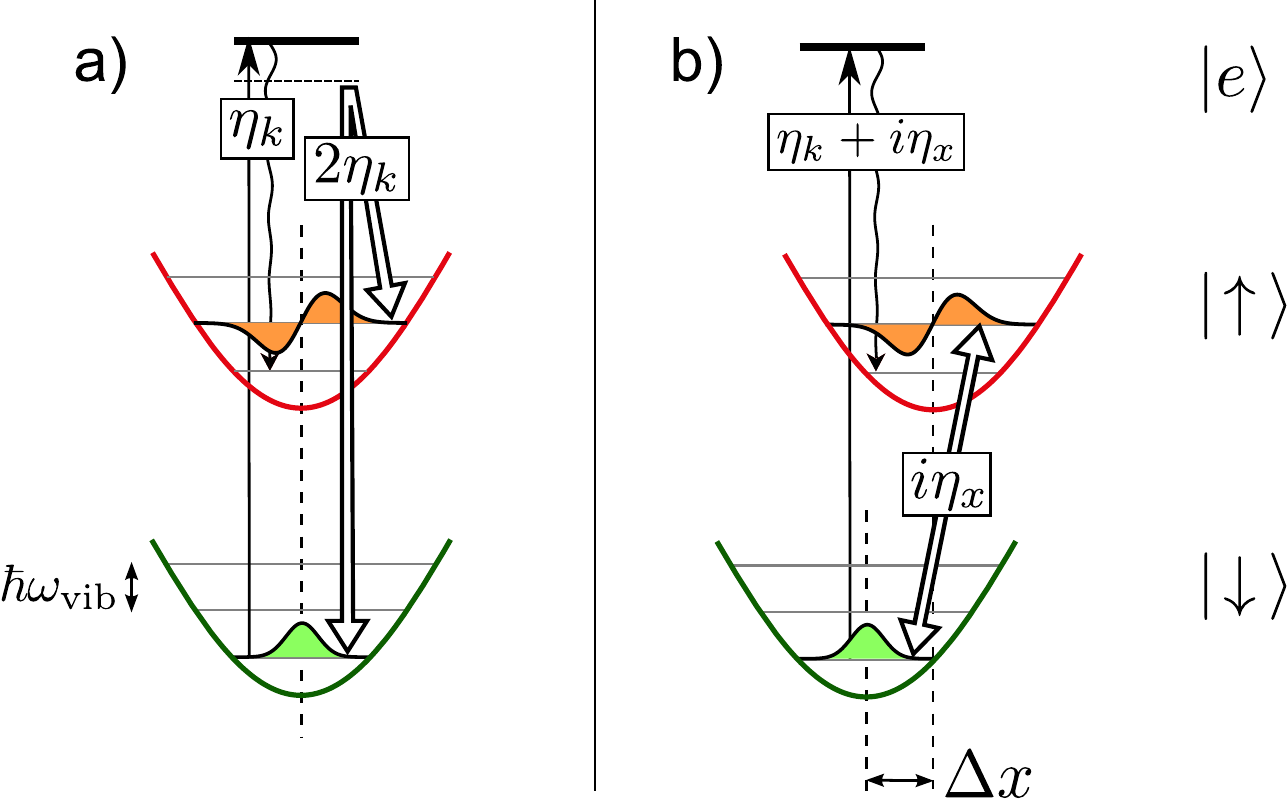}
\par\end{centering}

\centering{}\caption{(a) Raman sideband cooling scheme: a two-photon Raman transition between two identical trapping potentials reduces the vibrational state, $\ket{\uparrow,n}\rightarrow\ket{\downarrow,n-1} $. The wavefunction is shifted in momentum space by $\hbar\Delta k$. (b) Microwave sideband cooling
scheme: a microwave transition between two shifted trapping potentials reduces the vibrational state. Note that we use here a blue sideband transition to reduce the vibrational state, instead of the typical usage of a red sideband transition \cite{Perrin1998}.\label{fig:ToyModel}}
\end{figure}

In the standard Raman-based sideband cooling schemes \cite{jessen,Kerman2000}
the sideband is induced by a two-photon transition where the coupling
is given by the matrix element
\begin{eqnarray}
\Omega_{n-1,n}^{\text{{Raman}}}\, & = & \,\langle\downarrow,n-1\vert\hat{T}_{\Delta k}\vert\uparrow,n\rangle\times\Omega_{0}^\text{Raman},\label{eq:lambdickeparameter}
\end{eqnarray}
where $\hat{T}_{\Delta k}\equiv\exp(i\hspace{1pt}\hat{x}\Delta k)$ is the momentum
shift operator and $\Delta k\approx2k_{\text{{opt}}}$ is the wavevector
difference between the two optical photons for counterpropagating
beams. From here on, it is understood  that all transitions occur on the same site, $r=r'$. In the microwave-based scheme we neglect the microwave photon
recoil, and the sideband coupling corresponding to a lattice shift
$\Delta x$ between nearest neighboring sites is then given by 
\begin{equation}
\Omega_{n-1,n}\,=\,\langle\downarrow,n-1\vert\hat{T}_{\Delta x}\vert\uparrow,n\rangle\times\Omega_{0}.\label{eq:MWSBcoupling}
\end{equation}
Using the harmonic approximation, the Raman and microwave sideband
couplings can be expanded to the first order in the parameters $\eta_{k}=\hbar\hspace{1pt} k_{\text{{opt}}}/(2\hspace{1pt}p_{0})$
and $\eta_{x}=\Delta x/(2\hspace{1pt}x_{0})$, as shown in table (\ref{tab:Raman-vs-MW}),
where $p_{0}=\sqrt{m_{\text{{Cs}}}\hbar\omega_{\text{{vib}}}/2}$ and
$x_{0}=\sqrt{\hbar/(2m_{\text{{Cs}}}\omega_{\text{{vib}}})}$ are the
momentum and spatial rms width of the ground-state wavefunction, respectively
\cite{Stenholm}. From table (\ref{tab:Raman-vs-MW}) we can note
a clear duality between momentum and spatial shifts in the two sideband
cooling methods. The duality is better emphasized by using the general
complex Lamb-Dicke parameter
\begin{equation}
\eta=  \eta_{k}+i\eta_{x}=\hbar\hspace{1pt} k_{\text{{opt}}}/(2\hspace{1pt}p_{0})+i\Delta x/(2\hspace{1pt}x_{0})=k_{\text{{opt}}}\hspace{1pt}x_0+i\hspace{1pt}p_0\hspace{1pt}\Delta x/\hbar,\label{eq:genLDP}
\end{equation}
which accounts for both degrees of freedom via the momentum and spatial
Lamb-Dicke parameters, $\eta_{k}$ and $\eta_{x}$, respectively.
This generalized approach was introduced first in ion systems to describe
microwave-induced sidebands in the presence of spin-dependent forces
\cite{wunderlich2001}.

In the Raman-based cooling schemes with identical trapping potentials,
the spatial Lamb-Dicke parameter $\eta_{x}$ vanishes and the heating
comes from the recoil of the optical repumping photons, as depicted
in the figure \figref{fig:ToyModel}{a}. In the microwave-based scheme
however, the generalized Lamb-Dicke parameter is complex and the heating
is caused by a combination of recoil and projection heating.

The energy gained by an atom from recoil heating after one cycle results
from two recoils, one from absorption and one from spontaneous emission,
and is therefore given by
\begin{equation}
\Delta E_{\text{{rec}}}=2E_{\text{{R}}},\label{eq:DErec}
\end{equation}
where $E_{\text{{R}}}=\hbar^{2}k_{\text{{opt}}}^{2}/2m_{\text{{Cs}\,}}$
is the optical photon recoil energy \cite{Itano}. This quantity does
not depend on the details of the potentials but only on the atom's
properties, and it expresses the overall three-dimensional recoil
heating. 

\begin{table}[t]
\caption{Raman vs.\ microwave sideband cooling in the harmonic approximation.
The sideband couplings are first-order expansions in $\eta_{k}$ and
$\eta_{x}$ in the Lamb-Dicke regime defined by $|\eta|=|\eta_{k}+i\eta_{x}|\ll1$,
under the harmonic approximation.\label{tab:Raman-vs-MW}}

\centering{}%
\begin{tabular}{l>{\centering}p{3cm}c}
\toprule 
 & Raman & Microwave\tabularnewline
\midrule
\addlinespace
Sideband coupling strength $\Omega_{n-1,n}/\Omega_{0}$ & $i2\eta_{k}\sqrt{n}$  & $-\eta_{x}\sqrt{n}$\tabularnewline
\addlinespace
Recoil heating per cycle & $2\hbar\omega_{\text{{vib}}}\:\eta_{k}^{2}$ & $2\hbar\omega_{\text{{vib}}}\,\eta_{k}^{2}$\tabularnewline
\addlinespace
Projection heating per cycle & --- & $\hbar\omega_{\text{{vib}}}\,\eta_{x}^{2}$\tabularnewline
\addlinespace
Overall heating per cycle & $2\hbar\omega_{\text{{vib}}}\:\eta_{k}^{2}$ & $\hbar\omega_{\text{{vib}}}\,(\eta_{x}^{2}+2\eta_{k}^{2})$\tabularnewline
\bottomrule
\end{tabular}
\end{table}

In the shifted potentials shown in figure \figref{fig:ToyModel}{b},
in addition to the recoil heating, the atom's motional energy increases
on average by the projection heating energy $\Delta E_{\text{{proj}}}$.
This is due to the non-vanishing projection of the atom's initial
vibrational state $\Ket{\downarrow,n}$ onto the vibrational basis $\Ket{\uparrow,m}$
of the final spin state in the optical repumping process. In the harmonic
approximation, with $H_{\text{{ext}}}=\hbar\omega_{\text{{vib}}}(n+1/2)$,
and after adiabatic elimination of the excited state $\Ket{e}$, the
projection heating contribution for a relative shift $\Delta x$ can
be derived as
\begin{eqnarray}
\Delta E_{\text{{proj}}} & = & \hbar\omega_{\text{{vib}}}\sum_{m}(m-n)\left|\Braket{m|\hat{T}_{\Delta x}|n}\right|^{2}=\nonumber\\
 & = & \sum_{m}\Braket{m|\hat{[H}\hat{T}_{\Delta x}-\hat{T}_{\Delta x}\hat{H}]|n}\Braket{n|\hat{T}_{\Delta x}^{\dagger}|m}=\nonumber\\
 & = & \Braket{n|(\hat{H}_{\text{{ext}}}(\Delta x)-\hat{H}_{\text{{ext}}})|n},\label{eq:DEproj}
\end{eqnarray}
where we have introduced the shifted Hamiltonian
\begin{equation}
\hat{H}_{\text{{ext}}}(\Delta x)=\hat{T}_{\Delta x}^{\dagger}\hat{H}_{\text{{ext}}}\hat{T}_{\Delta x}=\frac{\hat{p}^{2}}{2m_{\text{{Cs}}}}+\frac{1}{2}m_{\text{{Cs}}}\omega_{\text{{vib}}}^{2}(\hat{x}+\Delta x)^{2}.
\end{equation}
The result of equation (\ref{eq:DEproj}) applies in general for any potential profile, and in the harmonic approximation it results in a quantity which is independent of $n$,
\begin{equation}
\Delta E_{\text{{proj}}}=\frac{1}{2}m_{\text{{Cs}}}\omega_{\text{{vib}}}^{2}\Delta x^{2},
\end{equation}
which is nothing but the potential energy difference as expected from
the semi-classical picture in figure (\ref{fig:classicalProj}). 

Using the same method, one can generally show that in the microwave
sideband cooling scheme the total average heating energy gained by
an atom in one cooling cycle is the sum of the recoil and projection
contributions. The total energy balance per cycle then becomes 
\begin{equation}
\Delta E_{\text{{tot}}}=\Delta E_{\text{{proj}}}+\Delta E_{\text{{rec}}}-\hbar\omega_{\text{{vib}}}=\hbar\omega_{\text{{vib}}}(\eta_{x}^{2}+2\eta_{k}^{2}-1).\label{eq:EnBalance}
\end{equation}
Similarly to the usual definition of the Lamb-Dicke regime \cite{LambDickeLimit}, the condition for cooling $\Delta E_{\text{{tot}}}<0$ defines a generalized Lamb-Dicke regime as the range where $|\eta|<1$.

\subsection{Quantitative model based on master equation}

The general theory of sideband cooling is very well known and has
been extensively studied in the literature \cite{wineland,Stenholm,Marzoli,Cirac92}.
Here, we discuss a quantitative model based on the Lindblad master
equation formalism. To provide a concrete example, we apply the model
to the level scheme of our specific system, though the model can be
adapted to other similar systems.

\begin{figure}[!b]
\begin{centering}
\includegraphics[scale=0.7]{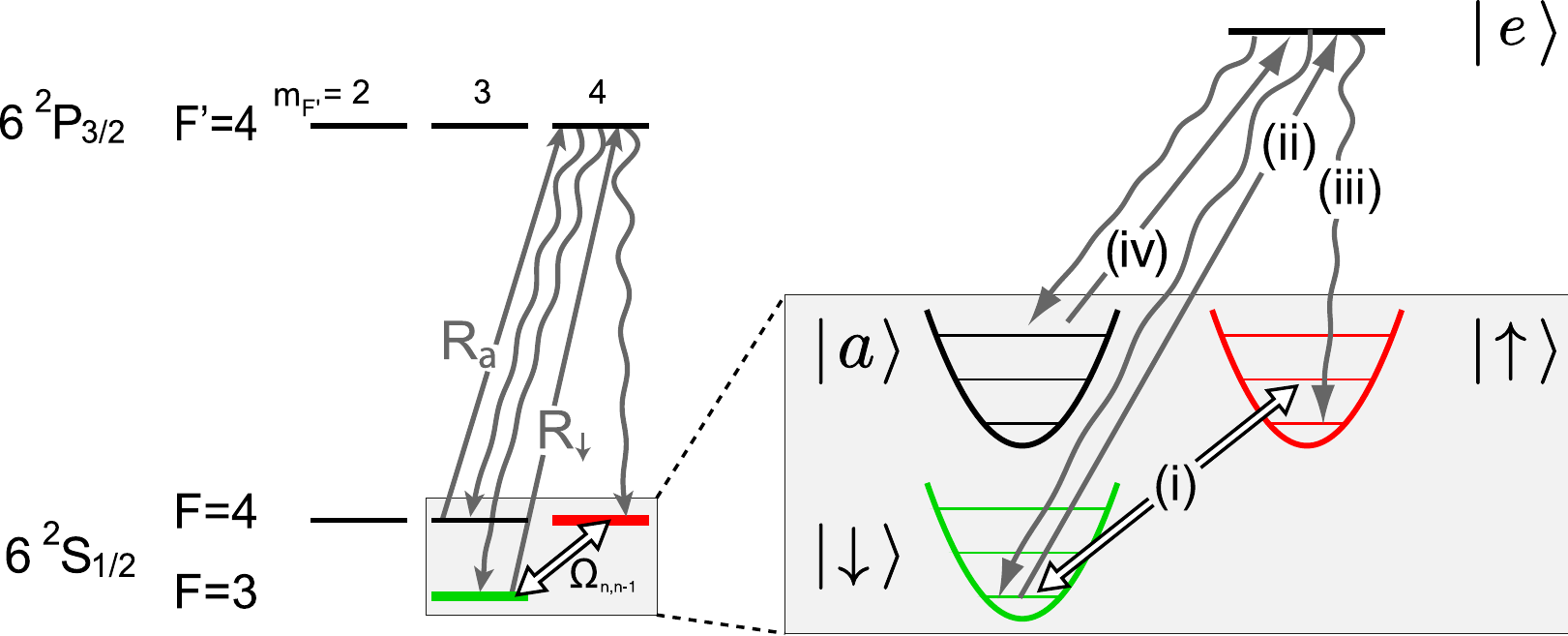}
\par\end{centering}

\caption{Microwave sideband cooling scheme in a realistic physical system using
$^{133}\text{Cs}$ atoms. (i) Microwave radiation tuned to the first
blue sideband induces a $\Ket{\uparrow,n}\rightarrow\Ket{\downarrow,n-1}$
transition decreasing the motional quantum number by one. (ii) The
cooling cycle is closed by an optical repumping transition $\Ket{\downarrow}\rightarrow\Ket{F'=4}$,
with rate $R_{\downarrow}$, and (iii) a spontaneous decay back to
state $\Ket{\uparrow}$. In (iv) an additional pumping laser brings
the atoms which have decayed to state $\Ket{a}$ back into the cooling
cycle, with rate $R_{a}$. Atoms reaching the dark state $\Ket{\uparrow,n=0\,}$
are out of the cooling cycle unless off-resonantly excited or heated
externally.\label{fig:cooling}}
\end{figure}

In the cooling cycle depicted in figure (\ref{fig:cooling}), microwave
radiation resonant with the first blue sideband transfers atoms from
states $\Ket{\uparrow,n}$ to states $\Ket{\downarrow,n-1}$. Concurrent
with the microwave, a $\sigma^{+}$-polarized repumper laser beam
couples state $\Ket{\downarrow}$ to state $\Ket{6^{2}P_{3/2},F'=4}\equiv\Ket{e}$,
from where the atoms close the cooling cycle by spontaneously decaying
back to state $\Ket{\uparrow}$. Due to the appreciable probability
of atoms decaying from state $\Ket{e}$ to state $\Ket{F=4,m_{F}=3}\equiv\Ket{a}$,
a second equally polarized pumping laser couples the two states and
brings the atoms which have decayed to state $\Ket{a}$ back into
the cooling cycle. In each cycle, an atom loses energy on average
until it reaches the ``dark state'' $\Ket{\uparrow,n=0}$ where
it is no longer affected by the microwave or the repumping lasers.
Nevertheless, a small probability remains that the dark state is depopulated
due to photon scattering from the lattice lasers or an off-resonant
microwave carrier transition.

To describe the cooling dynamics, we reduce the problem at hand to
an effective model with three spin states with the set of motional
states associated with each one of them. The considered Hilbert space
is then the one spanned by the states $\Ket{s,n}$, with $n$ being
the vibrational level and $\Ket{s}$ being one of the three internal
states $\Ket{\uparrow},$$\Ket{\downarrow}$ or $\Ket{a}$. The optically
excited state $\Ket{e}$ is adiabatically eliminated due to its very
short lifetime, $\tau=\SI{30}{\nano\second}$, compared to the motional
time scale. We use the Lindblad master equation formalism to write
the time evolution of the effective model's density matrix as \cite{Cirac92}
\begin{eqnarray}
\frac{d\rho}{dt} & = & -\frac{i}{\hbar}[\hat{H}'_{0}+\hat{H}_{\text{MW}},\rho]+\mathcal{L}[\rho],\label{eq:MasterEq}
\end{eqnarray}
where $\hat{H}_{0}'$ is the extension of the Hamiltonian from equation
(\ref{eq:H0}) to the states $\Ket{a,n}$
\begin{equation}
\hat{H}_{0}'=\sum_{s=\{\uparrow,\downarrow,a\}}\sum_{n}\varepsilon_{s,n}(\Delta x)\Ket{s,n,r}\Bra{s,n,r},\label{eq:H0'}
\end{equation}
and $\mathcal{L}$ is the Lindblad superoperator with the projectors
\begin{equation}
L_{n,r,s}^{n'\!,r'\!,s'}=\ket{s,n,r}\bra{s'\!,n'\!,r'},
\end{equation}
and the effective decay rates $\gamma_{s,n,r}^{s'\!,n'\!,r'}$ for the
transitions $\Ket{s'\!,n'\!,r'}\rightarrow\Ket{s,n,r}$ which are given
by

\begin{equation}
\gamma_{s,n,r}^{s'\!,n'\!,r'}=\alpha_{s}R_{s'}\left\langle |M_{s,n,r}^{s'\!,n'\!,r'}|^{2}\right\rangle _{\vec{k}_{\text{sp}}}\!\!,\label{eq:gamma}
\end{equation}
with
\begin{equation}
M_{s,n,r}^{s'\!,n'\!,r'}=\langle n,r,s|\,\hat{T}_{\Delta k_{s,s'}}\,\hat{T}_{\Delta x}|s'\!,n'\!,r'\rangle\,.
\end{equation}
Here, $\alpha_{s}$ is the branching ratio for the spontaneous emission
from state $\Ket{e}$ to state $\Ket{s}$, and $R_{a}$, $R_{\downarrow}$ are the pumping and repumping rate, respectively,
as shown in figure~(\ref{fig:cooling}). In addition, we account for the possibility that an atom in $\ket{\uparrow}$ state scatters a photon from the lattice with the rate  $R_{\uparrow}$. The matrix element
$M_{s,n,r}^{s'\!,n'\!,r'}$ accounts for the relative spatial shift between
the two involved vibrational states and for the transferred momentum
of both optical photons $\Delta k_{s,s'}=k_{\text{opt}}+\vec{k}_{\text{sp}}\cdot\vec{e}_{x}$
in the optical repumping process, with $\vec{k}_{\text{sp}}$ being
the wavevector of the spontaneously emitted photon and $\vec{e}_{x}$
being the unit vector along the lattice direction. Additionally, one
has to perform an average over $\vec{k}_{\text{sp}}$, indicated by
the angle brackets in equation (\ref{eq:gamma}) .

\begin{figure}[b]
\begin{centering}
\includegraphics[scale=0.4]{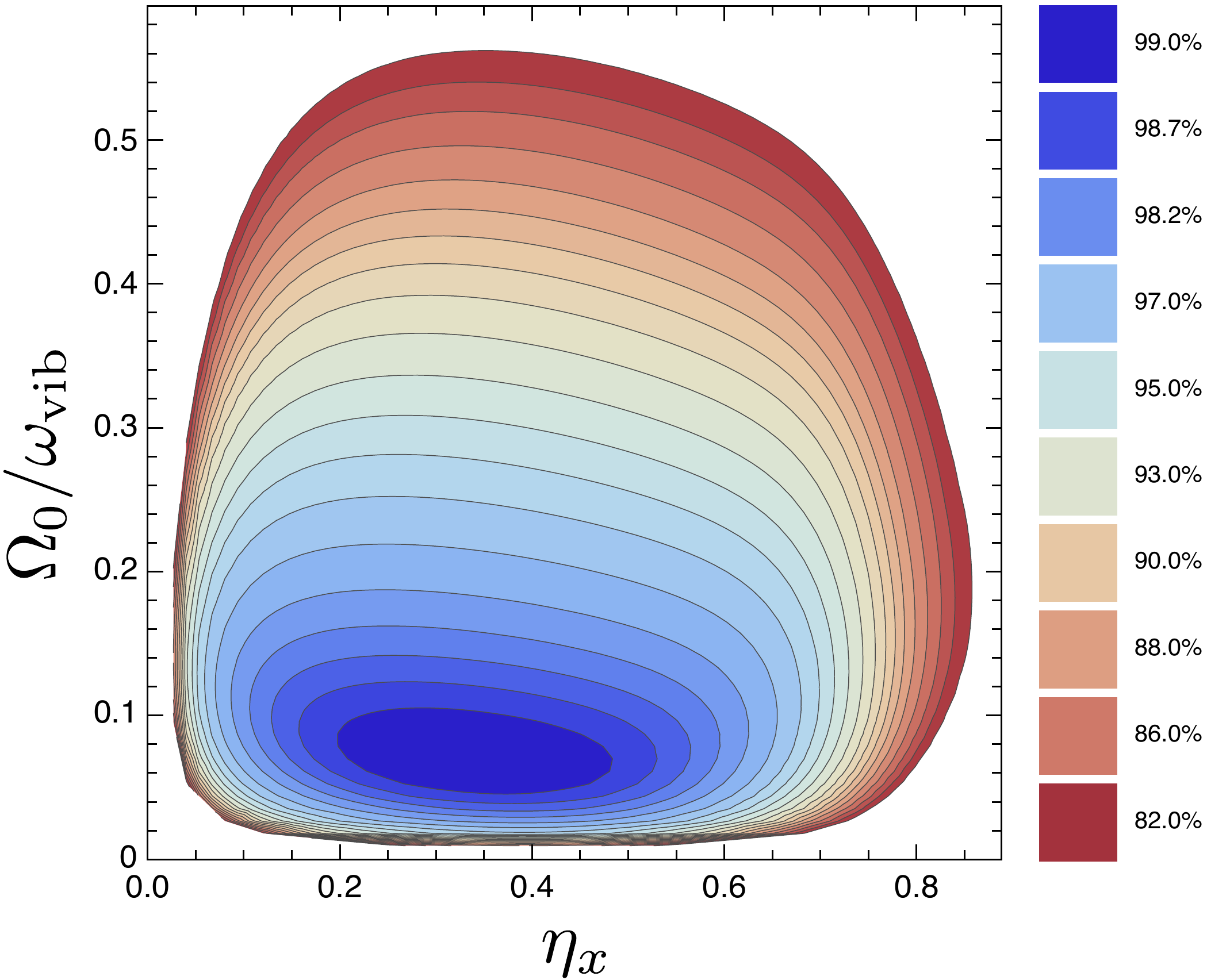}
\par\end{centering}

\centering{}\caption{Steady state population in the motional ground state $P_{\Ket{n=0}}$
as a function of $\eta_{x}$ and the bare microwave Rabi frequency $\Omega_{0}$.\label{fig:Cooling-simulation}}
\end{figure}

Given our experimental parameters, we compute the steady-state solution
to equation (\ref{eq:MasterEq}) numerically, using the same approximations
as in section \ref{sub:2.3}. In the computation, the microwave is
resonant with the $\Ket{\uparrow,1}\to\Ket{\downarrow,0}$ transition.
The rates $R_{a}$ and $R_{\downarrow}$ are set by the experimental values, which are chosen comparable to $\Omega_{0,1}$ and smaller than the vibrational level separations to avoid off-resonant transitions by power broadening of the vibrational levels of the $F=3$ ground state. Figure (\ref{fig:Cooling-simulation})
shows a contour plot of the ground state population $P_{\Ket{n=0}}\equiv\sum_{s,r}P_{\Ket{s,0,r}}$
in the steady state as a function of the bare microwave Rabi frequency
and the relative shift distance expressed in terms of $\eta_{x}$. 
When projection heating dominates, $\eta_{x}\gtrsim\eta_{k}$,
the energy balance in equation (\ref{eq:EnBalance}) just requires $\eta_{x}<1$ for cooling; for instance, figure~(\ref{fig:Cooling-simulation}) shows that a ground state population $P_{\Ket{0}}>80\%$ can be reached with $\eta_x<0.8$. For very small lattice shifts however,
with $\eta_{x}\ll1$, the microwave coupling for the blue sideband
transition becomes small compared to that of the carrier transition,
which renders the microwave action of removing an energy $\hbar\omega_{\text{vib}}$
per cycle inefficient compared to the recoil heating, which is the
dominant heating source for such small shifts. Weak microwave sideband
coupling and hence inefficient microwave cooling will also be present
at very low bare Rabi frequency, namely at the Rabi frequencies where
the sideband coupling becomes lower than the rate of depopulation
of the dark state. For high Rabi frequencies of the same order of
magnitude as the vibrational level spacing, the carrier coupling becomes
comparable to the blue sideband coupling, and the microwave cooling
action is again reduced.

\subsection{Experimental results\label{sub:3.3}}

\begin{figure}[b]
\begin{centering}
\hfill{}\includegraphics[scale=0.525]{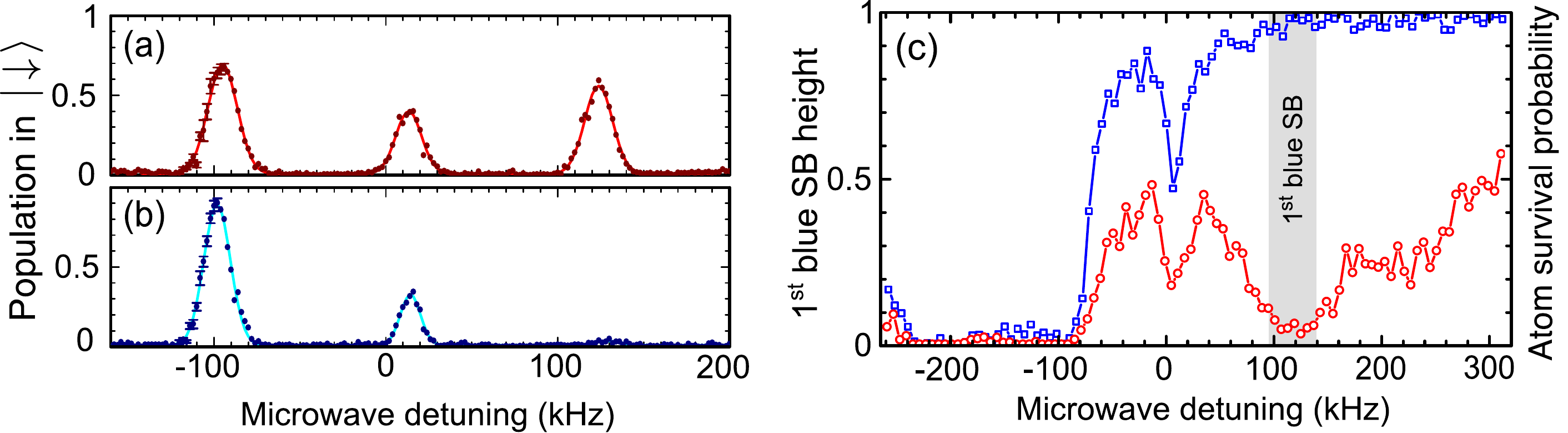}\hfill{}
\par\end{centering}

\centering{}\caption{Microwave spectroscopy performed (a) \label{fig:ScanningCoolingA} before cooling and (b) \label{fig:ScanningCoolingB}after
$\SI{20}{\milli\second}$ of microwave cooling, with optimal experimental
parameters (see Table \ref{tab:OptimCoolPar}). (c) \label{fig:ScanningCoolingC}Blue sideband
height vs.\;detuning of the microwave cooling frequency ({\small$\textcolor{red}\circ$})
and atom survival probability measured after the cooling ({\tiny$\textcolor{blue}\square$}), (data points in (a) and (b) are from \cite{Leonid}).\label{fig:ScanningCooling}}
\end{figure}

Microwave cooling is obtained by applying microwave radiation on resonance
with the first blue sideband, $\Ket{\uparrow,1}\to\Ket{\downarrow,0}$,
for a certain duration $\tau_{\text{cooling}}$, at a certain lattice
shift $\Delta x$, concurrently with the two optical pumping lasers
as shown in figure (\ref{fig:cooling}). In order to probe the final
vibrational state distribution, we record a spectrum of the first
order sideband transitions using a Gaussian microwave pulse satisfying
the $\pi$-pulse condition for the first red sideband, figure \figref{fig:ScanningCooling}{a}.
In the low temperature limit, the height of the first blue sideband
peak provides a good measure of the motional ground state population,
$P_{\Ket{\uparrow,0}}$, and thus of the cooling efficiency. For instance,
for atoms in the ground state one expects to detect no blue sideband.
Figures \figref{fig:ScanningCooling}{a} and \figref{fig:ScanningCooling}{b}
show two microwave spectra recorded before and after cooling, clearly
indicating a reduction of temperature by the cooling process.

In order to determine the optimum cooling parameters, the blue sideband
height is remeasured while scanning different variables, namely the
optical pumping intensities, the cooling microwave power and frequency,
the lattice shift distance and the duration of the cooling pulse.
Figure \figref{fig:ScanningCooling}{c} shows a scan of the cooling microwave
frequency. As indicated by a nearly zero detected signal from the
blue sideband, the optimum frequency for cooling lies evidently in
the vicinity of the first blue sideband, while a less pronounced cooling
is also present at the position of the second blue sideband. Furthermore,
the measurement reveals the absence of the blue sideband signal in
a broad range extending to negative detunings in addition to a weak
dip at the position of the carrier. These two observations are correlated
with a decrease in the atom survival given in the same figure. This
shows that, instead of being due to cooling, the absence of the blue
sideband here is due to increased atom losses. In fact, for zero and
negative microwave detunings, that is if the microwave is resonant
with the carrier $\Ket{\uparrow,n}\to\Ket{\downarrow,n}$ or red sideband
$\Ket{\uparrow,n}\to\Ket{\downarrow,n+m}$ transitions respectively,
the energy of the atom increases on average in each cooling cycle.
In the case of zero detuning the increase is due mainly to recoil
and projection heating in the absence of microwave cooling, while for
negative detunings microwave sideband heating occurs in addition to
the recoil and projection heating.

Once the optimum cooling parameters have been determined, we extract
the achieved steady state temperature assuming a thermal Boltzmann
distribution and neglecting the anharmonic spacing of the vibrational
levels. The ratio between the red and blue sideband heights is proportional
to the Boltzmann factor which is related to the average motional quantum
number $\langle n\rangle$ by
\begin{equation}
\frac{P_{\uparrow,1}}{P_{\uparrow,0}}=\exp(-\frac{\hbar\omega_{x}}{k_{B}T})=\frac{\langle n\rangle}{\langle n\rangle+1}\label{eq:CooledTemperature}
\end{equation}
Using the fitted sideband heights from figure \figref{fig:ScanningCooling}{b},
we calculate $\langle n\rangle=0.03\pm0.01$, and a ground state population
of $P_{\uparrow,0}\simeq97\%$, corresponding to a temperature $T\simeq\SI{1.6}{\micro\kelvin}$.
Table (\ref{tab:OptimCoolPar}) summarizes the optimum cooling parameters
for our setup.

\begin{table}
\begin{raggedright}
\caption{Optimal microwave cooling parameters for $\omega_{\text{{vib}}}=2\pi\times\SI{116}{\kilo\hertz}.$\label{tab:OptimCoolPar}}

\par\end{raggedright}

\centering{}%
\begin{tabular}{ccccccc}
\toprule 
$\Omega_{0}/2\pi$ & $\eta_{x}$ & $\eta_{k}$ & $R_{\downarrow}$ & $R_{a}$ & $R_\uparrow$ & $\tau_{\text{cooling}}$\tabularnewline
\midrule 
$\SI{16}{\kilo\hertz}$ & $0.3$ & $0.1$ & $\SI{35}{\milli\second^{-1}}$ & $\SI{35}{\milli\second^{-1}}$ & $\SI{15}{\second^{-1}}$ & $\SI{20}{\milli\second}$\tabularnewline
\bottomrule
\end{tabular}
\end{table}

\section{Motional state control}

\subsection{Motional state detection\label{sub:4.1}}

\begin{figure}[!h]
\begin{centering}
\includegraphics[scale=0.8]{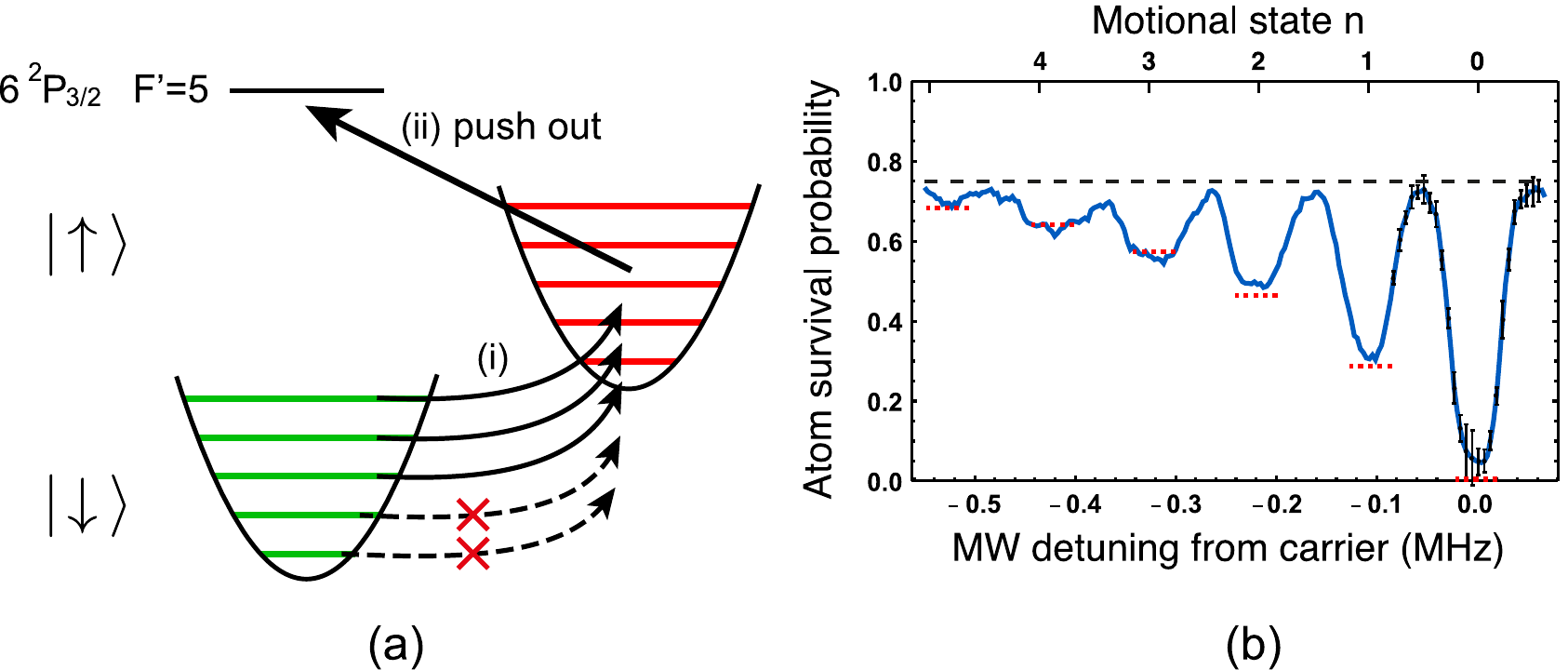}
\par\end{centering}
\centering{}\caption{(a) \label{fig:Filtering-schemeA}Method for measuring vibrational state population distributions:
(i) an initial microwave pulse resonant with the $n^{th}$ red sideband
transfers all atoms in states $\Ket{\downarrow,m}$, with $m\geqslant n$,
to state $\Ket{\uparrow}$; (ii) the transferred atoms are pushed
out of the lattice; (i) and (ii) are repeated $N$ times to overcome
low pulse efficiency. (b) \label{fig:Filtering-schemeB}Surviving fraction of atoms for a thermal
state, with the dotted lines indicating a thermal distribution of
$T\approx\SI{11.6}{\micro\kelvin}$; this temperature is compatible
with the independently measured one of about $\SI{10}{\micro\kelvin}$.
For each sideband $n$, after $N=3$ repetitions of the microwave
pulse plus push-out, only the atoms in states $m<n$ survive. The
horizontal dashed line indicates the maximum survival probability,
limited by the off-resonant transitions during the repeated pulses.
For the sake of clarity, error bars have
been displayed for the carrier transition only.\label{fig:Filtering-scheme}}
\end{figure}

We have developed a vibrational state detection scheme which allows
us to determine the vibrational state distribution of any given motional
state. It relies on removing all atoms above a selected vibrational
state $n$ from the trap and counting the remaining atoms, as illustrated
in figure \figref{fig:Filtering-scheme}{a}. The distribution is then
reconstructed from the differences of subsequent measurements. 

Atoms are first transferred to state $\Ket{\downarrow}$ by means of an adiabatic passage microwave pulse that is resonant with the carrier transition in unshifted lattices, which preserves vibrational states' populations. A microwave pulse resonant with the
red sideband $\Ket{\downarrow,n}\rightarrow\Ket{\uparrow,0}$ transfers
atoms from states $\Ket{\downarrow,m}$ with $m\geqslant n$ to states
$|\uparrow,m-n\rangle$. The transferred atoms are eventually pushed out
of the trap (see section \ref{sub:sec2.2}). However, since the sideband
transition rates depend on the initial vibrational state $\Ket{\downarrow,n}$
(due to, e.g., trap anharmonicity and Franck-Condon factor differences)
the microwave pulse does not achieve full transfer efficiency for
all transitions. To overcome this limitation, the procedure of microwave
pulse plus push-out is repeated several times to deplete all vibrational
states $\Ket{\downarrow,m}$ with $m\geqslant n$. If $f$ is the
population transfer efficiency for a given $n$, then after $N$ repetitions
the effective population transfer efficiency becomes $f'=1-(1-f)^{N}$.
For instance, an initial efficiency of $f=70\%$ is thus increased
to $f'\sim97\%$ with $N=3$ repetitions. Measuring the fraction of
remaining atoms as a function of the microwave frequency, we obtain
a sequence of plateaus at the successive sideband frequencies, as
shown in figure \figref{fig:Filtering-scheme}{b}. The plateau corresponding
to the $n^{th}$ sideband indicates the integrated population of states
$m<n$, that is, the cumulative distribution function $F_{n}=\sum_{m=0}^{n-1}p_{m}$,
from which the individual populations of the vibrational states are
then derived.

\subsection{Motional state engineering}

\begin{figure}
\begin{centering}
\includegraphics[width=11cm]{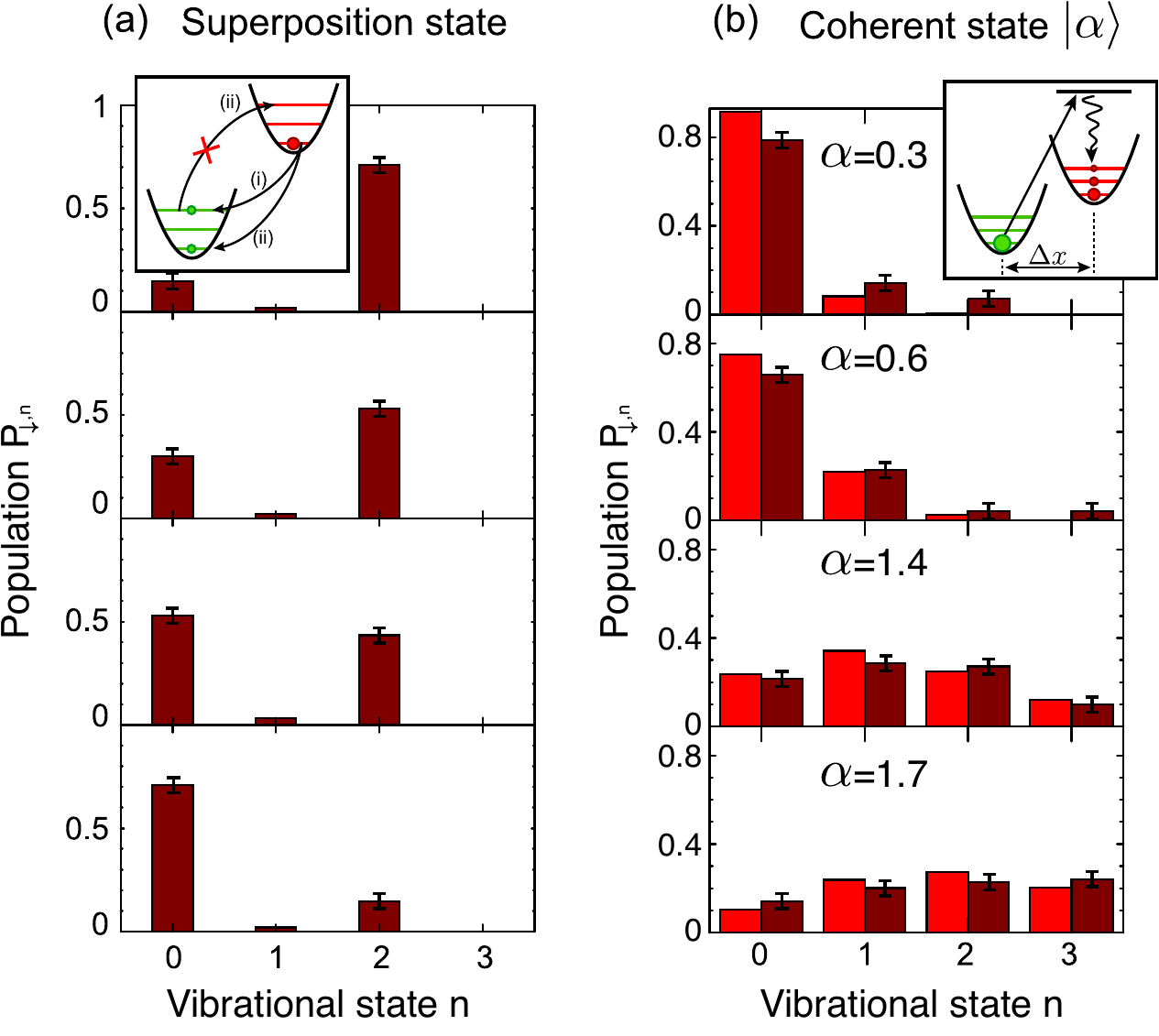}
\par\end{centering}

\caption{Motional state preparation and analysis. Shown are the populations
of the vibrational states $n=0,..,3$ after (a) creating superposition
states of $\Ket{n=0}$ and $\Ket{n=2}$ with different weights (from top to bottom, area of the first MW pulse 0.30, 0.40, 0.55, 0.70 in units of $\pi$) and (b) coherent vibrational states for different amplitudes $\alpha$, where the
left bars (brighter red) indicate the theoretically expected populations. The analysis technique used here, see figure~\figref{fig:Filtering-scheme}{a}, can only measure vibrational states' populations but not coherences.\label{fig:Motional-state-preparation}}
\end{figure}

With 97\% of the atoms cooled to state $\Ket{\uparrow,n=0}$ (see
section \ref{sub:3.3}), controlled preparation of different motional
states is possible using a combination of microwave pulses and selected
lattice shifts.

The simplest state that can be prepared is the Fock state $\Ket{\downarrow,m}$.
It requires addressing the $m$-th red sideband transition at the
lattice shift $\Delta x$ chosen to maximize the coupling $\Ket{\uparrow,0}\leftrightarrow\Ket{\downarrow,m}$.
The fidelity for preparing this state is limited by the cooling efficiency,
the population transfer efficiency and the selectivity of the microwave
pulse. Using an adiabatic passage pulse \cite{Khudaverdyan}, a state
preparation fidelity close to 98\% has been achieved for states up
to $m=6$.

A superposition of two Fock states is created by a two-pulse sequence
as shown in the inset of figure \figref{fig:Motional-state-preparation}{b}.
An initial microwave pulse resonant with the transition $\Ket{\uparrow,0}\to\Ket{\downarrow,2}$,
performed at the lattice shift which maximizes the coupling for the
transition, generates the state
\begin{equation}
\Ket{\psi}=c_{\uparrow,0}\Ket{\uparrow,0}+c_{\downarrow,2}\Ket{\downarrow,2}
\end{equation}
with variable coefficients $c_{\uparrow,0}$ and $c_{\downarrow,2}$
determined by the pulse duration. The lattice shift $\Delta x$ is
then changed to the distance at which the Franck-Condon factor for
the transition $\Ket{\uparrow,2}\leftrightarrow\Ket{\downarrow,2}$
is zero. The shifting is precisely timed so that the probability of
changing the vibrational state by the acceleration of the lattices
is zero \cite{Karski}. At the new lattice shift, a second microwave
pulse resonant with the carrier transition maps the population $|c_{\uparrow,0}|^{2}$
onto $\Ket{\downarrow,0}$. One would expect as a result a coherent
superposition between $\Ket{\downarrow,0}$ and $\Ket{\downarrow,2}$.
However, because of the appreciable duration of the sequence of $\SI{320}{\micro\second}$
(two sideband-resolved pulses plus lattice shift operation) compared
to the total spin coherence time of $\sim\SI{250}{\micro\second}$
in our setup, the coherence between the two vibrational states is
partially lost during the preparation procedure. This is a technical limitation which can be overcome by improving the coherence time, for instance in our setup, by cooling the transverse motion of the atoms to the three-dimensional ground state \cite{Kaufman2012}. This scheme represents a relevant step towards the use of the vibrational state as
the physical carrier for a qubit and/or the preparation of arbitrary
motional superposition states when working with neutral atoms.

In the same vein of engineering motional states, we project the state
$\Ket{\downarrow,n=0}$ onto a shifted potential to create the state
\begin{equation}\label{eq:coherentstate}
\Ket{\alpha}=\hat{T}_{\Delta x}\Ket{\uparrow,n=0}=e^{\alpha a^{\dagger}-\alpha^{*}a}\Ket{\uparrow,n=0}
\end{equation}
with $\alpha=\eta_{x}$. We realize this by applying an optical repumping
pulse while the lattice is displaced by $\Delta x$. This corresponds
to exciting the transition $\Ket{\downarrow}\to\Ket{e}$ followed
by a spontaneous decay to state $\Ket{\uparrow}$, which occur on
a time scale much shorter than the oscillation period of the atom
in the trap. We also neglect the recoil transferred by the optical
repumping photons, which is equivalent to assuming $\eta_{k}=0$.
Because the decay process additionally involves transitions to states $\ket{a}$ and $\ket{\downarrow}$, the resulting state is a statistical mixture of the three internal states; our analysis scheme however measures the vibrational population of the state projected on $\ket{\uparrow}$ state shown in equation (\ref{eq:coherentstate}). The statistical mixture can be avoided by replacing the optical repumping pulse and spontaneous decay by a fast two-photon Raman transition.
Measuring the population distribution of the created state reveals
a clear agreement with the theoretical expectation, as shown in figure
\figref{fig:Motional-state-preparation}{b}. With the state detection
scheme presented in section \ref{sub:4.1}, so far we can only measure
populations, while coherences could be accessed in the future through
interferometric schemes.

\section{Conclusions and outlook}
We have shown that microwave sideband transitions in spin-dependent optical lattices are a favorable alternative to Raman transitions for sideband cooling and motional state engineering. The effective Lamb-Dicke parameter can be continuously adjusted from zero to above one, giving the possibility to address directly higher-order sideband with coupling frequencies comparable to the bare Rabi frequency. We investigated the performance of microwave sideband cooling in the generalized Lamb-Dicke regime, and we compared it to the Raman sideband cooling; our analysis can be easily extended to the three-dimensional case~\cite{weiss2}.  

Quantum engineering of motional states represents one of the most attractive uses of microwave-induced sidebands. We demonstrated here a first step towards the creation of superposition between Fock states, and the preparation of coherent states. In the future, the interest resides in proving the coherence properties of these states through interferometric schemes, for instance, by  measuring the accumulated phase between two distinct Fock states, or through quantum beat experiments~\cite{Goto}.

Along the same line, spin-dependent shift operations can be employed to transfer a state-dependent momentum kick, allowing the realization of a superposition of opposite coherent states, producing Schr\"odinger-cat-like states as has been realized with ions~\cite{Monroe:1996}.

Microwave control of atomic motion in a spin-dependent optical lattice can be of interest for storing and processing quantum information via the motional states \cite{Haffner:2008}. For instance, the strength of coherent collisions for atoms close to the motional ground state exhibits a marked dependence on the relative motional state, which can be exploited, in analogy to \cite{Jaksch99}, to realize maximally entangled states in the motional degree of freedom.

In addition, microwave sideband transitions open the way for quantum transport experiments, where continuous tunneling between adjacent lattice sites occurs when $\Delta x$ is close to $d/2$, i.e., close to half the lattice spacing \cite{Mischuck:2010}.

Finally, it is worth noting that the microwave cooling technique studied here does not strictly require the use of the ``magic'' wavelength for the lattice potential, but
can still be operated with the same efficiency at other wavelengths, e.g., at $\lambda_{\text{L}}=\SI{1064}{\nano\meter}$ as we have tested. In fact, the optimal cooling efficiency occurring at around $\eta_x\sim 0.3$, see figure~(\ref{fig:Cooling-simulation}), can be reached by adjusting the polarization angle $\theta$.

% With such
% capabilities, all previous applications attempted with trapped ions
% are in principle now available also with strongly localized neutral
% atoms using microwaves, e.g. our techniques can be used in the verifications
% of fundamental quantum thermodynamics relations with single atoms
% \cite{Jarzynski}. 

\vspace*{-2mm}\ack{}{We thank Andreas Steffen and Tobias Kampschulte for fruitful discussions. Comments by an anonymous referee helped improve the manuscript. We acknowledge financial support by the DFG Research Unit (FOR 635), NRW-Nachwuchsforschergruppe ``Quantenkontrolle auf der Nanoskala'', AQUTE project, and  Studienstiftung des deutschen Volkes. AA acknowledges also support by Alexander von Humboldt Foundation.}

\vspace*{-2mm}\section*{References}{}
\bibliographystyle{biblio}
\bibliography{Bibliography}

\end{document}